\documentclass[prl,twocolumn,showpacs]{revtex4}
\usepackage{graphicx}
\usepackage{bm}
\usepackage{amsmath}

\newcommand{\pq}[2]{{#1}^{(\!#2\!\!\>)}}
\newcommand{\refeq}[1]{(\ref{#1})}
\newcommand{\reffig}[1]{FIG. \ref{#1}}
\renewcommand{\th}{\theta}
\newcommand{\defeq}{\stackrel{\mathrm{def}}{=}}

\begin{document}
\title{
Roundabout relaxation: collective excitation requires a detour to equilibrium
}
\author{Hidetoshi Morita}
\email{morita@complex.c.u-tokyo.ac.jp}
\author{Kunihiko Kaneko}
\affiliation{
Department of Basic Science,
Graduate School of Arts and Sciences,
The University of Tokyo,
3-8-1 Komaba, Meguro-ku, Tokyo 153-8902, Japan
}
\date{\today}

\begin{abstract}
Relaxation to equilibrium after strong and collective excitation is studied,
by using a Hamiltonian dynamical system of one dimensional XY model.
After an excitation of a domain of $K$ elements,
the excitation is concentrated to fewer elements,
which are made farther away from equilibrium, and
the excitation intensity increases logarithmically with $K$.
Equilibrium is reached only after taking this ``roundabout'' route,
with the time for relaxation diverging asymptotically as 
$K^\gamma$ with $\gamma \approx 4.2$.
\end{abstract}


\maketitle

Relaxation to equilibrium has been one of the most important topics
in non-equilibrium phenomena.
The relaxation dynamics after weak excitation has been thoroughly investigated;
it is represented as a superposition of the dynamics of excited modes,
each of which depends on external parameters such as temperature.
When several modes are strongly and collectively excited, on the other hand,
interaction among the excited modes is not negligible,
which may cause non-trivial dynamic behaviors
that are not simply represented by the superposition of the excited modes.
In particular, a set of the excited modes may form a partial system,
and give an internal state
that cannot be determined solely with an external conditions.
The relaxation will depend on the internal state, and, in turn,
the internal state will dynamically change with the relaxation;
such an interplay between the internal state and the relaxation
is expected to be seen in systems far from equilibrium in general.
We intend to search for some non-trivial and universal
relaxation phenomena therein.

Previously the authors have reported such a novel relaxation phenomenon,
using a Hamiltonian dynamical system of XY model
with mean field coupling~\cite{Morita-Kaneko};
when a part of the system is highly excited,
the relaxation progresses intermittently through bottlenecks
by self-organizing a critical state for the partial excited system.
Now it is interesting to study the relaxation
in a corresponding lattice system.
In particular we study a one dimensional (1D) XY model,
which has no phase transition and accordingly no critical state.
In spite of its absence, we find a rather remarkable relaxation process,
i.e. ``roundabout'' relaxation that a partially excited system
reaches equilibrium only after it once goes farther away from equilibrium.
In the present Letter, we report this discovery and analyze its mechanism,
by emphasizing the divergence of the relaxation time to equilibrium
with the number of excited elements.

The Hamiltonian that we study is~\cite{XY1D}
\begin{equation}
{\cal H}=\sum_{i=1}^{N}\left\{
\frac{p_i^2}{2} + J[1-\cos(\th_{i+1}-\th_i)] \right\},
\end{equation}
where $\th_i\in [-\pi,\pi)$,
with periodic boundary condition $\th_i=\th_{i+N}$.
We consider the case of $J>0$ (ferromagnetic),
setting $J=1$ without loss of generality.
The $N$ pendula, placed on the 1D lattice,
are coupled by the nearest neighbor interaction.
Each pendulum has two types of motion;
rotation at a higher and libration at a lower energy.

Note that when a single element is highly excited,
its relaxation is slow, with the relaxation time increasing
exponentially with the excitation momentum,
as in the case with mean field coupling~\cite{Nakagawa-Kaneko}.
This is simply because the effective interaction
between the excited element and the others
rapidly decreases as the excitation momentum increases.
The exponential dependence as such is discussed
in terms of Boltzmann-Jeans conjecture (BJC)~\cite{BJC}.

Here we study the relaxation when a domain of the system
consisting of quite a few elements is highly excited,
while the whole system is much larger~\cite{Morita-Kaneko}.
To be specific, the system is prepared to be in equilibrium,
with an energy density $U$ and total momentum zero.
At $t=0$, $K$ elements are simultaneously
excited with the same momentum $P_K$.
The momentum profile then becomes almost rectangular,
with two domains separated by two interfaces.
The number of the excited elements are much smaller
than that of the total system, $K/N\ll 1$;
we keep $K/N=0.1$ in the following numerical simulation.

\begin{figure}[tb]
\begin{center}
\includegraphics[scale=0.45]{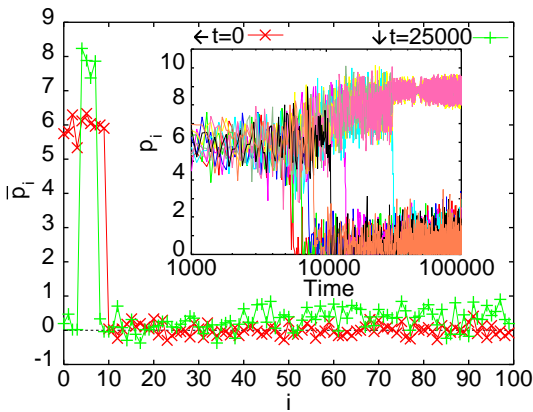}
\caption{
Two snapshots of the momentum profile $\overline{p}_i$,
with $\overline{\cdot}$ as the time average over a period of 1000,
at t=0 and 25000 (corresponding to the arrows to the inset).
In the inset, the corresponding time series of the momenta
of the excited elements are overlaid,
where abscissa axis is logarithmic scale in time.
$N=100, K=10, P_K=6, U=0.3.$
}
\label{fig:tseri}
\end{center}
\end{figure}

We first numerically observe the relaxation process.
The element at the interfaces of the excited domain is intermittently
absorbed into the not-excited domain, losing its excited energy.
The population of the excited elements $\pq{N}{E}$ accordingly
decreases one by one from the initial $K$ toward zero (equilibrium).
If the excitation is strong enough,
the excited domain receives some positive momentum from the escaped element,
to increase its center-of-mass momentum (CMM);
the process is repeated as the relaxation progresses (\reffig{fig:tseri}).
The excited elements then rotate faster and faster,
and thereby go farther away from equilibrium.
The increase of the CMM implies more time for the next escape,
as in the argument for BJC,
and requires much longer time to reach equilibrium.

Thus the highly excited state is self-sustained,
and the relaxation to equilibrium must take this ``roundabout'' route.
This process is not observed near equilibrium,
and requires sufficient excitation of a domain of elements.

Both of the two domains are almost in equilibrium
in the inertial frame of their own CMMs,
since their elements interact mostly among themselves.
The interaction at the interface, on the other hand, is much weaker,
because the difference between each CMM is much larger
and the interaction decreases rapidly with it.
Since both domains are composed of a sufficient number of elements,
each of which is regarded as a thermodynamic system,
and accordingly the whole system as the two thermodynamic systems
weakly coupled through the interface.

Being large enough, the non-excited system plays the role of a heat bath,
whose thermodynamic state is kept almost constant.
The state of the excited domain, on the other hand,
changes with relaxation, in the time scale
much slower than that for equilibration within the domain
but much faster than the whole relaxation to the equilibrium.
Thus it is possible to define an \textit{internal state}
for the excited system as its thermodynamic state.
To quantify the internal state,
we introduce the \textit{effective temperature} of the excited domain,
\begin{equation}
\pq{T}{E} \defeq \frac{1}{\pq{N}{E}} \pq{\sum_i}{E} (p_i-\pq{P}{E})^2,
\end{equation}
in the inertial frame with the following CMM of the excited domain,
\begin{equation}
\pq{P}{E} \defeq \frac{1}{\pq{N}{E}} \pq{\sum_i}{E} p_i,
\end{equation}
where $\pq{\sum}{E}$ denotes summation over only the excited elements.

\begin{figure}[tb]
\begin{center}
\includegraphics[scale=0.4]{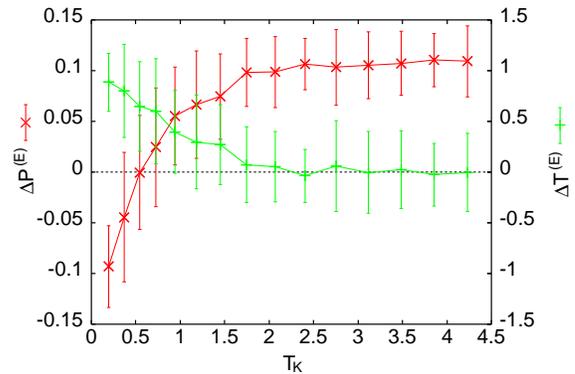}
\caption{
The change of
the center-of-mass momentum $\pq{\Delta P}{E}$ ($\times$, solid line)
and the effective temperature $\pq{\Delta T}{E}$ ($+$, dotted line),
which are time averaged over a period of 100,
against the (initial) effective temperature $T_K$.
$N=300, K=30, P_K=8$.
Obtained from the numerical calculations of 25 samples for each $T_K$.
}
\label{fig:dPdT_vs_T}
\end{center}
\end{figure}

To study the dynamics of the internal state,
we calculate the change of the CMM, $\pq{\Delta P}{E}$,
and of the effective temperature, $\pq{\Delta T}{E}$, of the excited domain,
due to the first absorption of an interfacial element
into the non-excited (bath) part ($K\to K-1$).
Their dependence on $T_K \defeq \pq{T}{E}|_{t=0}$
is plotted in \reffig{fig:dPdT_vs_T},
which is computed by preparing the excited state with $\pq{T}{K}$
from a thermal equilibrium state with the CMM $P_K$.
The figure shows that $\pq{\Delta T}{E}$ decreases
toward 0 from the positive side, indicating that
$\pq{T}{E}$ increases to approach a certain temperature.
On the other hand, $\pq{\Delta P}{E}$ approaches a certain positive value.
Except for the K dependence of the limit value of $\pq{\Delta P}{E}$,
the other properties above are independent of any other parameter
including $U$ and sufficiently large $P_K$.
By noting that
single elements are absorbed successively in the relaxation course,
the above result indicates that the effective temperature
increases and approaches a high value,
and there the CMM increases by some constant value per each absorption.
The CMM of the excited domain thus continues to increase,
until the last element is absorbed,
while a high temperature at the excited domain is sustained.

Now we study the increase of the relaxation time
with the number of the excited elements.
Numerically it is not so easy to trace all the relaxation course,
since it requires a huge time.
Instead we give an analytical estimate for it,
by focusing on the dynamics of the macroscopic quantities of the excited part.
To make notations simple in the following analysis,
subscripts $\pq{\cdot}{E}$ is omitted,
and let $\cdot_k$ be macroscopic quantities
when the number of the remaining excited elements is $k$.

\begin{figure}[bt]
\begin{center}
\includegraphics[scale=0.4]{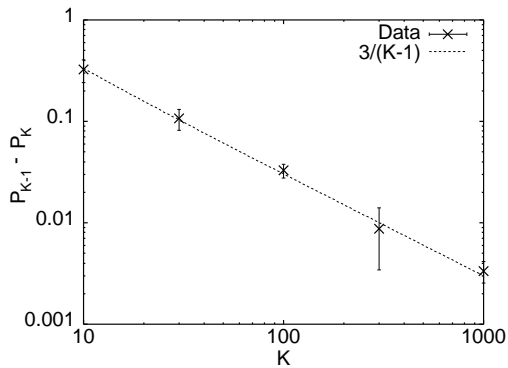}
\caption{$P_{K-1}-P_K$ versus $K$.
$P_K=8, T_K\approx 2.4$.
Obtained from the numerical calculations of 25 samples for each $K$.
}
\label{fig:dP_vs_K}
\end{center}
\end{figure}

Firstly we study how the change of the CMM $P_{k-1}-P_k$
depends on the population of the excited elements $k$.
$P_{K-1}-P_K$ with its dependence on $K$, obtained numerically,
is plotted in \reffig{fig:dP_vs_K}.
We have also confirmed that the relationship is almost not affected
by $P_K$ and $T_K$, as long as they are large enough.
Thus we obtain, 
\begin{equation}
P_{k-1}-P_k=\alpha/(k-1).
\label{eq:dP_vs_k}
\end{equation}

Here $\alpha$ is a constant with the dimension of momentum,
which we numerically found $\alpha \approx 3$.
We examine a rough estimation of the value,
by considering  the dynamics of only three interfacial elements:
an escaping element with two neighboring ones.
Furthermore the motion of each neighboring element is approximated
as a constant rotation with the momentum 0 and $\Omega$.
Then the effective Hamiltonian of the interfacial dynamics can be given as,
\begin{equation}
{\cal H}_{inter}=p^2/2+J[1-\cos\th]+J[1-\cos(\th-\Omega t)].
\end{equation}
For $|\Omega|> \Omega_{thr}\approx 6$,
two motions withing the region around $p\sim 0$ and $p\sim\Omega$
are separated by KAM tori,
while for $|\Omega|< \Omega_{thr}$,
an orbit can cross between the two regions,
due to the collapse of the last KAM torus
through the so-called ``resonance overlap~\cite{Chirikov}.''
In other wards,
the interface element can jump over from $p\sim\Omega$ to $p\sim 0$,
only if $|\Omega|$ is less than $\Omega_{thr}$.
Coming back to the original problem,
this is nothing but the escape of the interfacial element
from the excited domain.
The loss of the momentum $\Omega_{thr}$ thereby
should be compensated by the the momentum gain of the neighboring elements,
each of which is given approximately $\Omega_{thr}/2$ from the symmetry.
Hence the excited domain gains momentum about $\Omega_{thr}/2$,
which is distributed to the $k-1$ elements therein,
leading to the increase of the CMM by $\Omega_{thr}/2/(k-1) \approx 3/(k-1)$.

\begin{figure}[bt]
\begin{center}
\includegraphics[scale=0.4]{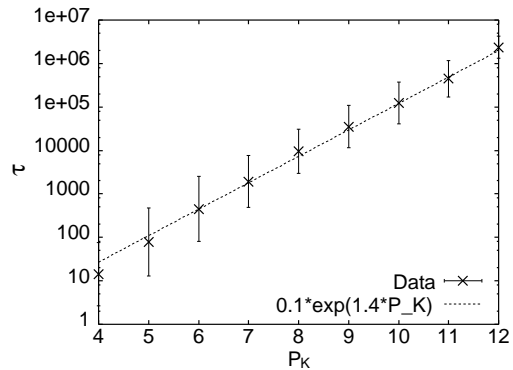}
\caption{$\tau$ versus $P_K$.
$N=300, K=30, T_K\approx 2.4$.
Obtained from the numerical calculations of 25 samples for each $P_K$.
}
\label{fig:tau_vs_P}
\end{center}
\end{figure}

Secondly we compute the relaxation time $\tau$
when the first one of the excited elements loses 
its energy and is absorbed into the non-excited part ($K\to K-1$),
whose dependence on $P_K$ is plotted in \reffig{fig:tau_vs_P}.
This exponential dependence is just the form of the BJC mentioned above.
This relationship again is almost not affected by $K$ and $T_K$~\cite{comment},
as long as they are large enough.
Thus we obtain,
\begin{equation}
\tau(P)=Ce^{\beta P}.
\label{eq:tau_vs_P}
\end{equation}
The constant $\beta$ has the dimension of inverse momentum,
which is the order of the inverse of the momentum
for the separatrix motion of a pendulum.
Numerically we have obtained $\beta\approx 1.4$ and $C\approx 0.1$.

On the basis of the two short-time properties
\refeq{eq:dP_vs_k} and \refeq{eq:tau_vs_P},
we can estimate the long-term relaxation.
First we consider the moment of the last excited element $P_1$.
When the population falls to $k$ from initial $K$,
the gain of CMM is,
\begin{eqnarray}
P_k-P_K &=& \sum_{j=k}^{K-1}(P_j-P_{j+1}) = \sum_{j=k}^{K-1}\frac{\alpha}{j}
\nonumber
\\
&=&\alpha(S_{K-1}-S_{k-1}),
\label{eq:Pk}
\end{eqnarray}
by summing up \refeq{eq:dP_vs_k}, where
\begin{equation}
S_K\defeq\sum_{j=1}^{K}\frac{1}{j},
\end{equation}
which increases asymptotically as $\log K$ with $K\to\infty$.
The momentum of the last excited element $P_1$ is then evaluated as,
\begin{equation}
P_1-P_K=\alpha S_{K-1} \to \log K.
\end{equation}
Hence the CMM diverges asymptotically as $\log K$.

Finally we estimate the total relaxation time to the equilibrium.
From \refeq{eq:tau_vs_P} and \refeq{eq:Pk},
the total time up to the relaxation from $k$ to $k-1$ is given by,
\begin{equation}
t_k=\sum_{j=k}^{K}\tau(P_j)
=t_K e^{\gamma S_{K-1}} \sum_{j=k}^{K} e^{-\gamma S_{j-1}}
\label{eq:t_k}
\end{equation}
where $\gamma\defeq\alpha\beta$ is a dimensionless number.
Then the relaxation time to equilibrium is estimated as,
\begin{equation}
\tau_{eq}=t_1
=t_K e^{\gamma S_{K-1}} \left(
1+\sum_{k=1}^{K-1} e^{-\gamma S_k}
\right).
\label{eq:tau_eq}
\end{equation}
Recalling $\log(k+1) < S_k \leq \log k+1$, for $k\geq 1$, we get
\begin{equation}
e^{-\gamma} k^{-\gamma}
\leq e^{-\gamma S_k}
< (k+1)^{-\gamma}.
\label{eq:exp_Sk_ineq}
\end{equation}
Combining \refeq{eq:tau_eq} and \refeq{eq:exp_Sk_ineq} yields
\begin{equation}
1 + \sum_{k=1}^{K-1} \frac{e^{-\gamma}}{k^\gamma}
\leq \frac{\tau_{eq}}{t_K} e^{-\gamma S_{K-1}}
< 1 + \sum_{k=1}^{K-1} \frac{1}{(k+1)^\gamma}.
\end{equation}
As $K\to\infty$, $e^{\gamma S_K}\to e^{\gamma\log K}=K^\gamma$, and
\begin{equation}
\sum_{k=1}^{K} \frac{1}{k^\gamma} \to
\begin{cases}
\zeta(\gamma) &\mbox{if $\gamma >1$}
\\
\log K &\mbox{if $\gamma=1$}
\\
K^{1-\gamma} &\mbox{if $0\le\gamma <1$},
\end{cases}
\end{equation}
where $\zeta(\cdot)$ is Riemann zeta.
Hence the asymptotic form of
the relaxation time to equilibrium $\tau_{eq}$ as $K\to\infty$ is
\begin{equation}
\tau_{eq} \to
\begin{cases}
K^\gamma &\mbox{if $\gamma >1$}
\\
K\log K &\mbox{if $\gamma=1$}
\\
K &\mbox{if $0\le\gamma <1$}.
\label{eq:tau_eq_asympt}
\end{cases}
\end{equation}

Note that, if each excited elements relaxes independently, $\tau_{eq}=K t_K$.
Thus $\gamma=1$ is the lower bound of the divergence
attributed to the cooperative effect of the excited elements.
Indeed, in the present model, we have
$\alpha\approx 3$ and $\beta\approx 1.4$, and we get $\gamma\approx 4.2$.
Hence the relaxation time satisfies $\tau_{eq}\to K^{4.2}$,
showing the rapid divergence with the number of excited elements.

\begin{figure}[tb]
\begin{center}
\includegraphics[scale=0.4]{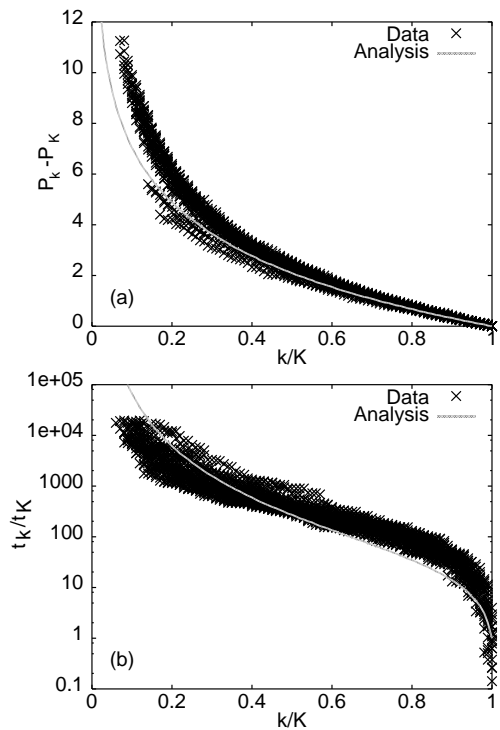}
\caption{
Temporal change of (a) $P_k$ and (b) $t_k$, plotted against $k/K$,
which decreases through the relaxation.
Numerical results from 25 samples of direct relaxation simulations
are plotted ($\times$),
while the solid curves are the analytical estimation
from \refeq{eq:Pk} and \refeq{eq:t_k}, respectively.
$N=1000, K=100, P_K=8, T_K\approx 4.2$.
}
\label{fig:P_and_t_vs_k}
\end{center}
\end{figure}

To check the validity of the above analysis,
we have computed the increase of CMM and the relaxation time
up to the $k$-th element, $P_k$ and $t_k$
(\reffig{fig:P_and_t_vs_k}-(a) and (b), respectively).
Even though it is hard to follow all the relaxation course numerically,
the numerical results (crosses) plotted against $k/K$
rather well agree with the theoretical estimates 
\refeq{eq:Pk} and \refeq{eq:t_k} (solid curves).

In summary,
we have discovered that a class of one-dimensional lattice system
must take a ``roundabout relaxation route,'' once highly excited.
After an excitation of a domain of the system,
the momentum of the domain increases logarithmically,
to go farther away from equilibrium, before reaching equilibrium.
The relaxation time accordingly diverges as a power $K^\gamma$
against the increase of the excited elements $K$.
Hence the relaxation with collective excitation has a rather peculiar form,
and is rather different from that near equilibrium.
Note that the present choice of one-dimensional XY (pendulum) model
is not so special; actually preliminary results
in two-dimensional case also suggest similar roundabout relaxation.
In physics there should be a variety of examples
described by coupled pendula on a lattice,
and our results will be relevant to slow relaxation
in various real non-equilibrium systems.
It is also notable that in a closed chemical reaction-diffusion system,
a self-sustained excited state is formed
as a transient dissipative structure~\cite{Awazu-Kaneko},
which leads to hindrance of relaxation to equilibrium.

The authors are grateful to A. Awazu and S. Honjo for discussion.
This work was supported by a Grant-in-Aid for Scientific Research
from the Ministry of Education, Science, and Culture of Japan.


\begin{thebibliography}{99}
\bibitem{comment}
There could be slight dependence on $T_K$.
However, since $\pq{T}{E}$ approaches a constant value,
this dependence influences only slightly on the exponent
in Eq.\refeq{eq:tau_eq_asympt}.
\end{thebibliography}
\end{document}